\documentclass[letterpaper,titlepage,11pt]{article}
\usepackage{latexsym,amssymb,amstext,amsmath}
\usepackage{slashed}
\usepackage{mathrsfs}
\usepackage{color}
\usepackage{enumerate}
\usepackage{graphicx}
\usepackage{tikz}
\usepackage{etex}
\usepackage{latexsym}
\usepackage{cite}

\usepackage[utf8]{inputenc}
\allowdisplaybreaks

\usepackage[
colorlinks=true,
linkcolor=blue,
urlcolor=red,
filecolor=green,
citecolor=red,
pdfstartview=FitV,
pdftitle={},
pdfauthor={Mehmet Ozkan},
pdfsubject={},
pdfkeywords={},
pdfpagemode=None,
bookmarksopen=true
]{hyperref}

\newcommand{\bea}{\setlength\arraycolsep{2pt} \begin{eqnarray}}
	\newcommand{\eea}{\end{eqnarray}}
\newcommand{\nn}{\nonumber}

\usepackage{hyperref}

\newsavebox{\uuunit}
\sbox{\uuunit}
{\setlength{\unitlength}{0.825em}
	\begin{picture}(0.6,0.7)
		\thinlines
		\put(0,0){\line(1,0){0.5}}
		\put(0.15,0){\line(0,1){0.7}}
		\put(0.35,0){\line(0,1){0.8}}
		\multiput(0.3,0.8)(-0.04,-0.02){12}{\rule{0.5pt}{0.5pt}}
		\end {picture}}

	\def\be{\begin{equation}}
		\def\ee{\end{equation}}
	\def\ba{\begin{array}}
		\def\ea{\end{array}}
	\def\bea{\begin{eqnarray}}
		\def\eea{\end{eqnarray}}
	\def\bd{\begin{displaymath}}
		\def\ed{\end{displaymath}}
	
	\def\nn{\nonumber}
	
	\def\a{\alpha}
	\def\b{\beta}
	\def\g{\gamma}
	
	\def\d{\delta}
	
	\def\e{\epsilon}

	\def\p{\psi}

	\def\l{\lambda}
	\def\L{\Lambda}
	\def\m{\mu}
	\def\n{\nu}
	\def\r{\rho}

	\def\t{\tau}

	\def\o{\omega}
	\def\O{\Omega}

	\def\nn{\nonumber}
	
	\def\cN{\mathcal{N}}

	\DeclareMathOperator{\diag}{diag}

	\makeatletter
	\@addtoreset{equation}{section}
	\makeatother
	
\begin{document}
		%
		\begin{titlepage}
			
			\bigskip
			
			\begin{center}
				{\LARGE\bfseries Carrollian and  Non-relativistic Jackiw-Teitelboim  Supergravity}
		
				\vspace{10mm}
				
				\textbf{Lucrezia Ravera}$^{\star,\ddagger}$,  \textbf{Utku Zorba}$^{\ast}$\\[5mm]
				\vskip 25pt
				%
				%
				{\em   \hskip -.1truecm $^\star$DISAT, Politecnico di Torino, Corso Duca degli Abruzzi 24, 10129 Torino, Italy. \vskip 5pt }
				{\em   \hskip -.1truecm $^\ddagger$INFN, Sezione di Torino, Via P. Giuria 1, 10125 Torino, Italy. \vskip 5pt }
				{\em   \hskip -.1truecm $^\ast$Physics Department , Boğaziçi University, 34342 Bebek, Istanbul, Turkey.  \vskip 5pt }
				{e-mail: {\tt lucrezia.ravera@polito.it}, {\tt utku.zorba@boun.edu.tr}}
				
			\end{center}
			
			\vspace{3ex}

			\begin{center}
				{\bfseries Abstract}
			\end{center}
			\begin{quotation}
				
				We present non- and ultra-relativistic Jackiw-Teitelboim (JT) supergravity as metric BF theories based on the extended Newton-Hooke and extended AdS Carroll superalgebras in two spacetime dimensions, respectively. The extended Newton-Hooke structure, and, in particular, the invariant metric necessary for the BF construction of non-relativistic JT supergravity, is obtained by performing an expansion of the $\mathcal{N}=2$ AdS$_2$ superalgebra. Subsequently, we introduce the extended AdS$_2$ Carroll superalgebra, and the associated invariant metric, as a suitable redefinition of the extended Newton-Hooke superalgebra. The mapping involved can be seen as the supersymmetric extension of the duality existing at the purely bosonic level between the extended Newton-Hooke algebra with (positive) negative cosmological constant and the extended (A)dS Carroll algebra in two dimensions. Finally, we provide the Carrollian JT supergravity action in the BF formalism. Moreover, we show that both the non-relativistic and the ultra-relativistic theories presented can also be obtained by direct expansion of $\mathcal{N}=2$ JT supergravity.

			\end{quotation}
			
			\vfill
			
			\flushleft{\today}
		\end{titlepage}
		\setcounter{page}{1}
		\tableofcontents
		
		\par\noindent\rule{\textwidth}{0.4pt}

		\section{Introduction}{\label{Intro}}
		
		Progresses in the understanding of the quantum nature of black holes have been enabled by Jackiw-Teitelboim (JT) gravity \cite{Teitelboim1983,Jackiw1984} (which is a model of $2D$ dilaton gravity, cf., e.g., \cite{Grumiller:2002nm,Grumiller:2021cwg}) and its holographic dual, especially concerning  quantum chaos and scrambling, which are indeed believed to be related to the black hole information paradox (for a comprehensive review see, e.g., \cite{Almheiri2,Bousso,Trunin:2020vwy}). 
        Besides, from the gauge/gravity duality perspective, JT gravity, as a two-dimensional toy model of quantum gravity describing universality in nearly extremal black holes \cite{Almheiri,Jensen}, is dual to the Sachdev-Ye-Kitaev (SYK) model \cite{SYK1,SYK2} at large $N$ and low energies. The SYK model is an exactly solvable quantum theory of $N >> 1$ interacting Majorana fermions (cf. \cite{Trunin:2020vwy,Sarosi2017} and references therein) that has received a growing attention since its formulation from both the high energy and the condensed matter physics communities, and it is nowadays considered one of the most notable models for quantum chaos and holography. 
        This is due to the fact that it exhibits some remarkable properties. For instance, it is exactly solvable in the large $N$ and IR limit, where, moreover, it acquires conformal symmetry and the effective action can be approximated by the Schwarzian one. Furthermore, it exhibits a holographic relation with JT gravity, with the latter describing excitations above the near-horizon extremal black hole and, once considered on the clipped Poincaré disk, effectively reducing to the one-dimensional theory with Schwarzian action.
        In particular, the SYK model resulted to be an excellent toy model for many physical phenomena, including quantum chaos \cite{Maldacena:2015waa}, information scrambling \cite{Sekino:2008he,Susskind:2011ap,Lashkari:2011yi}, traversable wormholes \cite{Maldacena:2017axo,Maldacena:2019ufo}, and strange metals \cite{Sachdev:2015efa}.
        
        On the other hand, another remarkable feature of JT gravity is that it can be regarded as a topological theory in two spacetime dimensions within the BF formalism, which is reminiscent of the three-dimensional formulation of Chern-Simons theory.\footnote{For recent developments and reviews of the three-dimensional Chern-Simons formulation of (super)gravity theories, in particular both non- and ultra-relativistic, see, e.g., \cite{Papageorgiou:2009zc,Bergshoeff:2016lwr,Hartong:2016yrf,Aviles:2018jzw,Ozdemir:2019orp,Ozdemir:2019tby,Ravera:2019ize,Concha:2019dqs,Ali:2019jjp,Concha:2019mxx,Kasikci2020,Concha:2020ebl,Concha:2020tqx,Concha:2020eam,Concha:2021jos,Concha:2021jnn,Concha:2021llq} and references therein.} 
        The BF formulation of JT gravity \cite{Fukuyama,Isler,Chamseddine} is particularly well-suited to consider other types of theories by extending the underlying symmetries. These symmetries can be either relativistic or ultra/non-relativistic (super)symmetries. Recently, the non-relativistic ($c\rightarrow \infty$, where $c$ is the speed of light) and ultra-relativistic ($c \rightarrow 0$, also called Carrollian, see, e.g., \cite{Bergshoeff:2017btm,Bergshoeff:2014jla,Duval:2014uoa,Duval:2014uva,Duval:2014lpa,Hartong:2015xda}) limits of JT gravity have been obtained in both the second-order and the BF formalism in \cite{Gomis2020,Grumiller2020}. Moreover, the BF setup allows one to consider the boundary theory of JT supergravity \cite{Grumiller2018}, the latter resulting to be the gravity dual of the supersymmetric extension of SYK model in the low energy limit \cite{Grumiller2018,Fu}. 
		
		Despite all the aforementioned rather recent developments and the promising applications of JT (super)gravity in various research fields, its non- and ultra-relativistic counterparts still remain little explored. 
		Conversely, non-relativistic theories have received a growing interest in the last years due to their relation to condensed matter systems \cite{Son:2008ye,Balasubramanian:2008dm,Kachru:2008yh,Christensen:2013lma,Christensen:2013rfa,Hartong:2014oma,Hartong:2014pma,Hartong:2015wxa} and non-relativistic effective field theories \cite{Hoyos:2011ez,Abanov:2014ula,Geracie:2015dea,Gromov:2015fda}, while Carroll symmetries appear, for instance, in high energy physics in the study of tachyon condensation \cite{Gibbons:2002tv}, warped conformal field theories \cite{Hofman:2014loa}, and tensionless (super)strings \cite{Bagchi:2013bga,Bagchi:2015nca,Bagchi:2016yyf,Bagchi:2017cte,Bagchi:2018wsn}.
		
		In the above discussed context, at least to our knowledge, consistent non- and ultra-relativistic JT supergravity theories have not been developed yet, especially within the BF formulation.  
		In this work, we present non- and ultra-relativistic JT supergravities as metric BF theories based, respectively, on the extended Newton-Hooke and extended AdS Carroll superalgebras in two spacetime dimensions. The former superalgebra, together with the associated invariant metric necessary to implement the metric BF construction, is obtained by performing a Lie algebra expansion of the $\mathcal{N}=2$ AdS$_2$ superalgebra. Consequently, we introduce the extended AdS$_2$ Carroll superalgebra and its invariant metric through a suitable redefinition of the extended Newton-Hooke superalgebra. Hence, we construct the Carrollian JT supergravity action in the BF formalism. Moreover, we show that both the non- and ultra-relativistic JT supergravity actions presented in this work can also be obtained by directly expanding JT supergravity formulated as a BF theory based on the $\mathcal{N}=2$ AdS$_2$ superalgebra.
		
		The paper is organized as follows: We start by briefly reviewing, in Sections \ref{JTgrBF} and \ref{N2JTsugra} respectively, the first-order formulation of JT gravity as a BF theory based on the AdS$_2$ algebra and the extension to the case of $\mathcal{N}=2$ JT supergravity as a BF theory based on the $\mathcal{N}=2$ AdS$_2$ superalgebra. In Section \ref{NRthy} we introduce the extended Newton-Hooke superalgebra as a Lie algebra expansion of the $\mathcal{N}=2$ AdS$_2$ one, deriving the associated invariant metric which allows us to consequently develop the non-relativistic JT supergravity action as a BF model. Subsequently, in Section \ref{URthy} we move on to the construction of the ultra-relativistic counterpart of JT supergravity. Hence, we derive the supersymmetric extended AdS$_2$ Carroll algebra, which can be obtained both as a redefinition of the extended Newton-Hooke superalgebra and as an expansion of the $\mathcal{N}=2$ AdS$_2$ one. The same applies at the level of the invariant metric, which permits us to construct the Carrollian JT supergravity theory in the BF setup. Section \ref{concl} is devoted to concluding remarks and possible future developments of our analysis. In Appendix \ref{appA} we collect our notation and conventions.

		\section{Review of Jackiw-Teitelboim gravity as a metric BF theory}\label{JTgrBF}
		
		In this section we will consider the first-order formulation of JT gravity as a BF theory based on the AdS$_2$ algebra \cite{Fukuyama, Isler, Chamseddine}. We will follow \cite{Grumiller2020}  and briefly review the structure of (metric) BF theories for later purposes.
		
		The BF theory action is given by
		\begin{equation}
		    \mathcal{S}^{\text{BF}} [\mathcal{X}^*,A] = \frac{k}{2\pi} \int_{\mathcal{M}_2} \mathcal{L}^{\text{BF}} [\mathcal{X}^*,A] \,, \label{actionBF1}
		\end{equation}
		with 
		\begin{equation}
		    \mathcal{L}^{\text{BF}} [\mathcal{X}^*,A] = \mathcal{X}^* F = X_K \left( dA^K + \frac{1}{2} C_{I J}{}^K A^I \wedge A^J  \right) \,,
		\end{equation}
		where $k$ is a dimensionless constant, $\mathcal{X}^*=X_I \mathrm{E}^I$ is a scalar transforming in the coadjoint representation of the Lie algebra $\mathfrak{g}$ on which the theory is based, and the Lie algebra valued 1-form $A^I_\mu \mathrm{e}_I dx^\mu$ is a gauge field with curvature two-form $F\equiv dA + \frac{1}{2}[A,A]$.
		The structure constants $C_{I J}{}^K $ of $\mathfrak{g}$ are defined by
		\bea
		[\mathrm{e}_I,\mathrm{e}_J] = C_{I J}{}^K \mathrm{e}_K \,, \label{liealgebra}
		\eea
		where $\mathrm{e}_I$ are the generators of $\mathfrak{g}$. The dual $\mathfrak{g}^*$ has basis $\mathrm{E}^I$ obeying $\mathrm{E}^I(\mathrm{e}_J)=\delta^I_J$. 
		
		The action \eqref{actionBF1} is invariant under gauge transformations
		\bea
		\delta_\l A^I = d \l^I  +  C_{J K}{}^I \l^K A^J   \,, \quad  \delta_\l X_I =  - C_{I J}{}^K \l^J X_K\,.\label{bfgauge}
		\eea
		The equations of motions of the theory read
		\bea
		F^I = d A^I + \frac{1}{2} C_{ J K}{}^I A^J \wedge A^K=0\,, \quad   d X_I +  C_{I J}{}^K A^J   X_K=0\,. \label{eombf}
		\eea
		
		If $\mathfrak{g}$ admits an \textit{invariant metric} $\langle \cdot , \cdot \rangle : \mathfrak{g} \times \mathfrak{g} \rightarrow \mathbb{R}$ that is a non-degenerate, symmetric, ad-invariant
        bilinear form,\footnote{By ad-invariance we intend $\langle [z,x],y \rangle = \langle x,[z,y] \rangle=0$ for all Lie algebra elements $x,y,z \in \mathfrak{g}$.} then one can use this metric to identify elements of the dual $\mathfrak{g}^*$ with elements of the Lie algebra $\mathfrak{g}$ by means of $\langle \mathcal{X} , \cdot \rangle = \mathcal{X}^* (\cdot)$, that is $X^I=g^{IJ}X_J$ (where $g_{IJ}=\langle \mathrm{e}_I, \mathrm{e}_J \rangle$ and $g^{IJ}g_{JK}=\delta^I_K$).
		
		In BF theories having an invariant metric is not a requirement, but in this paper we are interested in a particular subclass of BF theories (see \cite{Grumiller2020}), called \textit{metric BF theories}, whose construction is in fact based on Lie algebras exhibiting an invariant metric.
		The action for a metric BF theory is given by
		\bea
		\begin{aligned}
		\mathcal{S}^{\text{mBF}}[\mathcal{X},A] & = \frac{k}{2\pi} \int_{\mathcal{M}_2} \mathcal{L}^{\text{mBF}}[\mathcal{X},A] = \frac{k}{2\pi} \int_{\mathcal{M}_2} \langle \mathcal{X},F \rangle \\
		& = \frac{k}{2\pi} \int_{\mathcal{M}_2} g_{L K} X^L \left(d A^K + \frac{1}{2} C_{I J}{}^K A^I \wedge A^J\right)\,, \label{bfaction}
		\end{aligned}
		\eea
		and its equations of motion read
		\begin{equation}
		    F=0 \,, \quad d \mathcal{X} + [A,\mathcal{X}] = 0 \,.
		\end{equation}
		A standard example of metric BF theories can be obtained by considering as a starting point simple Lie algebras where one can use the matrix trace to write $\mathcal{L}^{\text{mBF}}=\text{tr}(\mathcal{X}F)$.
		
		The first-order formulation of (AdS) JT gravity as a metric BF theory was developed by considering the AdS$_2$ algebra, which is spanned by the Lorentz and spacetime translations generators $\lbrace{J,P_A\rbrace}$, with $A,B,\ldots=0,1$, and read 
		\begin{align}
			\left[ J, P_{A} \right] &= \epsilon_{AB} P^B\,, & \left[ P_{A}, P_{B} \right] &= - \L \epsilon_{AB} J \,, \label{adsbosonic}
		\end{align}
		where $\L= -\frac{1}{\ell^2}$ is the cosmological constant written in terms of the AdS$_2$ radius $\ell$, and $\e_{01}=-\e_{10} = 1$. This algebra admits an invariant metric given by
		\bea
		&& \langle P_A , P_B \rangle = \frac{\eta_{AB}}{\ell^2} \,, \quad \langle J , J \rangle = 1\,. \label{bosonicmetric}
		\eea
		In this setup, the gauge field and the coadjoint scalar take the following form (cf. \cite{Grumiller2020}):
		\bea
		A = E^A P_A + \O J\,, \quad \mathcal{X} =  X^A P_A + X J\,, \label{gaugefields}
		\eea
		and the corresponding covariant curvatures are given by 
		\bea
		R(E)^A &=& d E^A - \epsilon^{AB} \O \wedge E_B \,,  \nn\\
		R(\Omega) &=& d\O  + \frac{1}{2 \ell^2}  \epsilon_{AB} E^A \wedge E^B  \,. \label{bosoniccurvatures}
		\eea
		Using \eqref{gaugefields}, \eqref{bosoniccurvatures}, and the invariant metric \eqref{bosonicmetric}, one obtains the BF theory formulation of JT gravity, that is
		\bea
		\mathcal{S}_{\text{JT}} = \frac{k}{2\pi} \int \left( - \L X^A R(E)_A + X R(\Omega) \right)\,. \label{bfJT}
		\eea
		The equations of motion of the theory are given by 
		\bea
		&&d E^A - \epsilon^{AB} \O \wedge E_B  = 0\,,  \nn\\
		&&d\O  + \frac{1}{2 \ell^2}  \epsilon_{AB} E^A \wedge E^B = 0 \,, \nn \\
		&&d X^A  - \epsilon^{AB} \O X_B  +  \epsilon^{AB} X E_B = 0\,, \nn \\
		&&dX - \frac{1}{\ell^2} \epsilon^{AB}X_A E_B = 0\,. \label{eom}
		\eea 
		In particular, the equations of motion of the fields $X^A$ enforce the two-dimensional torsion constraint for the zweibein $E^A$. Upon solving the latter for the spin connection $\Omega$ and plugging it back into the action, one is left with the second-order action for JT gravity, where $X$ turns out to be the dilaton field.
		In this way, one can derive the JT equations of motion in the second-order formalism (see \cite{Gomis2020}), 
		\bea
		&& \mathcal{R} - 2 \L = 0\,,\nn \\
		&&\nabla_\m  \nabla_\n X - \L g_{\m\n} X = 0\,, \label{jteom}
		\eea
		$\mathcal{R}$ being the curvature scalar.  
		The first of these equations yields locally AdS$_2$ spacetime, and the second can be considered as a back-reaction of the field $X$ in response to the metric $g_{\mu \nu}$.
		
		\section{$\mathcal{N}=2$  Jackiw-Teitelboim supergravity}\label{N2JTsugra}
		
		In this section, we will review the first-order formulation of $\mathcal{N}=2$ JT supergravity as a BF theory presented in  \cite{Grumiller2018}, adopting, however, the slightly different notation and conventions of \cite{Freedman2012} (see also Appendix \ref{appA}). 
		In order to proceed with the construction of the theory, it is convenient to introduce the $\mathcal{N}=2$ AdS$_2$ superalgebra as a starting point.\footnote{We will be interested in studying the non-relativistic counterpart of JT supergravity, which results to be achievable and well-defined in particular once we start from the $\mathcal{N}=2$ relativistic theory. This is analogous to what was done, for instance, in  \cite{Bergshoeff:2016lwr} for the three-dimensional Chern-Simons formulation of extended Bargmann gravity.}
		The $\mathcal{N}=2$ AdS$_2$ superalgebra is given by the commutation relations
		\begin{align}
			\left[ J, P_{A} \right] &= \epsilon_{AB} P^B\,, & \left[ P_{A}, P_{B} \right] &= \frac{1}{\ell^2}\epsilon_{AB} J \,, \nn\\
			[J, Q^i ] &=  \frac{1}{2} \g_{*} Q^i\,, & [P_{A}, Q^i ] &= \frac{1}{2 \ell} \g_{A}  Q^i \,, \nn \\
			[U, Q^i ] &=  - \frac{1}{2\ell} \varepsilon^{ij}  Q^{j}\,,\label{ads2susy1}
		\end{align}
		and the anticommutator
		\begin{equation}
		    \{ Q^i_\a  ,  Q^j_\b  \} =  \delta^{ij} (\g_A C^{-1})_{\a \b} P^A  + \frac{\delta^{ij}}{\ell} (\g_* C^{-1})_{\a \b} J - \varepsilon^{ij} ( C^{-1})_{\a \b}  U \,, \label{ads2susy2}
		\end{equation}
		where $C$ is the charge conjugate matrix, defined as  $C= i \sigma_2$, $\L = -\frac{1}{\ell^2}$ is the cosmological, $i,j=1,2$ label the number of supercharges, and $\varepsilon_{ij}$ is the two-dimensional Levi Civita symbol ($\varepsilon_{12}=-\varepsilon_{21}=1$). The superalgebra \eqref{ads2susy1}-\eqref{ads2susy2} is endowed with an invariant metric
		\bea
		&&\langle P_A,P_B \rangle = \frac{\eta_{AB}}{\ell^2} \,, \quad \langle J,J \rangle = 1\,, \quad \langle U,U \rangle = \frac{1}{\ell^{2}} \,,\nn \\
		&& \langle Q^i_\a,Q^j_\b \rangle = \frac{2}{\ell} \delta^{ij} (C^{-1})_{\a\b}\,. \label{invar}
		\eea
		The gauge connection 1-form, the coadjoint scalar, and the curvature 2-form associated with the $\mathcal{N}=2$ AdS$_2$ superalgebra are, respectively,\footnote{In the following, we will frequently omit the spinor index $\alpha$ to lighten the notation.}
		\bea
		A &=& E^A P_A + \O J +  T U  + \bar{\Psi}^i Q^i \,, \nn \\
		\mathcal{X} &=& X^A P_A + X J + Y U + \bar{\l}^i Q_i \,,  \nn \\
		R&=& R(E)^A P_A + R(\Omega)  J +  R(T) U  + R(\Psi)^i Q^i \,, \label{gauge}
		\eea
		where now
		\bea
		R(E)^A &=& d E^A - \epsilon^{AB} \O \wedge E_B + \frac{1}{2} \delta_{ij}\bar{\Psi}^i \g^A \wedge \Psi^j  \,,  \nn\\
		R(\Omega) &=& d\O  + \frac{1}{2 \ell^2}  \epsilon_{AB} E^A \wedge E^B + \frac{1}{2 \ell} \delta_{ij}\bar{\Psi}^i \g_* \wedge \Psi^j \,, \nn \\
		R(T) &=& d T  - \frac{1}{2} \varepsilon_{ij} \bar{\Psi}^i  \wedge \Psi^j\,, \nn \\
		R(\Psi)^i &=& d \Psi ^i + \frac{1}{2} \O \wedge \g_* \Psi^i  + \frac{1}{2\ell} E^A \wedge \g_A \Psi^i  + \frac{1}{2 \ell} \varepsilon^{ij} T \wedge  \Psi^j\,. \label{curvatures}
		\eea
		Notice that the $\lambda^i$ in \eqref{gauge} are fermionic, while $X^A,X,Y$ are bosonic.
		Our convention for the gamma matrices in two dimensions is given in Appendix \ref{appA}.
		
		We can then construct the BF theory of JT supergravity by using \eqref{gauge} and \eqref{curvatures} along with the invariant metric \eqref{invar}. The resulting action takes the following form: 
		\bea
		\mathcal{S}_{\text{sJT}} = \frac{k}{2\pi} \int \left( \frac{1}{\ell^2}X^A R(E)_A + X R(\O) + \frac{1}{\ell^{2}}Y  R(T) +  \frac{2}{\ell}\bar{\l}^i R(\Psi)_i \right)\,. \label{sugraJT}
		\eea
		One can show that the action \eqref{sugraJT} is invariant under the following supersymmetry transformations:  
		\bea
		\d E^A &=& - \delta_{ij} \bar{\e}^i \g^A \Psi^j\,,\nn \\
		\d \O &=& - \frac{1}{\ell} \delta_{ij} \bar{\e}^i \g_* \Psi^j\,,\nn \\
		\d T &=& \varepsilon _{ij} \bar{\e}^i \Psi^j\,,\nn \\
		\d \Psi^i &=& d \epsilon^i  + \frac{1}{2} \g_* \e^i \O  +  \frac{1}{2 \ell} \g_A \e^i E^A + \frac{1}{2 \ell} \varepsilon^{ij} \e^j T\,, \nn \\
		\d X^A &=& - \delta_{ij} \bar{\e}^i \g^A \l^j \,,\nn\\
		\d X &=& - \frac{1}{\ell} \delta_{ij} \bar{\e}^i \g_* \l^j \,,\nn\\
		\d Y &=&  \varepsilon_{ij} \bar{\e}^i \l^j \,,\nn\\
		\d \l^i &=&  \frac{1}{2\ell} \g_A X^A \e^i + \frac{1}{2} \g_* X \e^i + \frac{1}{2\ell} \varepsilon^{ij} Y \e^j\,, \label{susytransformation}
		\eea
		where $\e^{i \alpha}$ are the supersymmetry parameters. 
		The equations of motion of the theory are
		\bea
		&& d E^A - \epsilon^{AB} \O \wedge E_B + \frac{1}{2} \delta_{ij}\bar{\Psi}^i \g^A \wedge \Psi^j = 0 \,,  \nn\\
		&& d\O  + \frac{1}{2 \ell^2}  \epsilon_{AB} E^A \wedge E^B + \frac{1}{2 \ell} \delta_{ij}\bar{\Psi}^i \g_* \wedge \Psi^j = 0 \,, \nn \\
		&& d T  -  \frac{1}{2} \varepsilon_{ij} \bar{\Psi}^i  \wedge \Psi^j = 0 \,, \nn \\
		&& d \Psi ^i + \frac{1}{2} \O \wedge \g_* \Psi^i  + \frac{1}{2\ell} E^A \wedge \g_A \Psi^i  + \frac{1}{2 \ell}\varepsilon^{ij} T \wedge  \Psi^j = 0 \,, \label{eomsusyAdS1}
		\eea
		that is $R(E)^A=R(\Omega)=R(T)=R(\Psi)^i=0$, coming from variation of the action with respect to $X^A,X,Y,\lambda^i$, respectively, together with
		\bea
		&&d X^A  - \epsilon^{AB} \O X_B  +  \epsilon^{AB} X E_B + \delta_{ij} \bar{\Psi}^i \gamma^A \lambda^j = 0\,, \nn \\
		&&dX - \frac{1}{\ell^2} \epsilon^{AB}X_A E_B + \frac{1}{\ell} \delta_{ij} \bar{\Psi}^i \gamma_\star \lambda^j = 0\,, \nn \\
		&& dY - \varepsilon_{ij} \bar{\Psi^i} \lambda^j = 0 \,, \nn \\
		&& d \lambda^i + \frac{1}{2\ell} \gamma_A E^A \lambda^i + \frac{1}{2}  \gamma_\star \Omega \lambda^i + \frac{1}{2\ell}  \varepsilon^{ij} T \lambda^j -\frac{1}{2\ell}  \gamma_A X^A \Psi^i - \frac{1}{2} \gamma_\star X \Psi^i - \frac{1}{2\ell} \varepsilon^{ij} Y \Psi^j = 0 \,. \label{eomsusyAdS2}
		\eea 
		The construction above is the first-order formulation of $\mathcal{N}=2$ JT supergravity. Let us move on to the study of its non- and ultra-relativistic counterparts.
		
		\section{Supersymmetric  extended Newton-Hooke algebra sNH$_2$ and non-relativistic Jackiw-Teitelboim supergravity}\label{NRthy}
		
		We will now develop the non-relativistic counterpart of the JT supergravity theory reviewed in the previous section.
		The theory will be based on the extended Newton-Hooke superalgebra with negative cosmological constant (which we name sNH$_2$) we are going to present in the following.
		
		\subsection{Extended Newton-Hooke superalgebra sNH$_2$}
		
		The extended Newton-Hooke superalgebra in two dimensions can be obtained by performing a Lie algebra expansion on the $\cN=2$ AdS$_2$ superalgebra. For a comprehensive treatment of the Lie algebra expansion procedure see, e.g., \cite{LAE2, LAE3, LAE4} and references therein.
		The supersymmetric extension of the (purely bosonic) extended Newton-Hooke (NH) algebra of \cite{Gomis2020} requires three fermionic generators, $Q^\pm_\a$ and $R_\a$. Besides, it requires two extra bosonic generators $U_1$ and $U_2$, both of them acting non-trivially on the fermionic generators in the presence of a cosmological constant $\Lambda$ (while they become central charges in the limit $\Lambda \rightarrow 0$, that is $\ell \rightarrow \infty$). In particular, the extra bosonic generators allows us to end up with a well-defined invariant metric.
		
		Before implementing Lie algebra expansion on \eqref{ads2susy1}-\eqref{ads2susy2}, we split the Lorentz indices as $A = (0,1)$, such that
		\bea
		P_A = \left( P_0, P_1\right)\,, \quad \gamma^A = \left(\gamma^0 , \gamma^1 \right) \,. \label{split}
		\eea
		In this way, one obtains the following decomposition of the AdS$_2$ superalgebra:
			\begin{align}
			\left[ P_0, J\right]  &= -  P_1\,,          &  \left[J, P_1 \right] &= P_0 \,,          &  \left[ P_0, 
			P_1 \right] &= \frac{1}{\ell^2} J\,, \nonumber\\
			[J, Q^i] &=  \frac{1}{2} \g_* Q^i \,,  &[P_0, Q^i] &=  \frac{1}{2\ell} \g_0 Q^i\,, &[P_1, Q^i] &=  \frac{1}{2\ell} \g_1 Q^i \,, \nn \\
			 [U, Q^i] &= - \frac{1}{2 \ell} \varepsilon^{ij} Q_j\,, 
			\label{adsdecom1}
		\end{align}
		together with
		\bea
		\{ Q^1_\a ,  Q^1_\b  \} &=& - (\g_0 C^{-1})_{\a \b} P_0 + (\g_1 C^{-1})_{\a \b} P_1   +  \frac{1}{\ell}   (\g_* C^{-1})_{\a \b} J \,,\nonumber\\
		\{ Q^2_\a ,  Q^2_\b  \} &=& - (\g_0 C^{-1})_{\a \b} P_0 + (\g_1 C^{-1})_{\a \b} P_1   +  \frac{1}{\ell}   (\g_* C^{-1})_{\a \b} J \,,\nonumber\\
		\{ Q^1_\a ,  Q^2_\b  \} &=&  - (C^{-1})_{\a \b} U \,.
		\label{adsdecom2}
		\eea
		In addition, we introduce the following combination of the fermionic generators $Q^1_\alpha$ and $Q^2_\alpha$:
		\bea
		\tilde{Q}^\pm = \frac{1}{\sqrt{2}} \left( Q^1 \pm \g_0 Q^2\right)\,. \label{fermionicdef}
		\eea
		Hence, the decomposed algebra takes the form
	    \begin{align}
			\left[ P_0, J\right]  &= -  P_1\,,          &  \left[J, P_1 \right] &= P_0 \,,          &  \left[ P_0, 
			P_1 \right] &= \frac{1}{\ell^2} J\,, \nonumber\\
			[J, \tilde{Q}^{\pm}] &=  \frac{1}{2} \g_*  \tilde{Q}^{\mp} \,,  &[P_0, \tilde{Q}^{\pm}] &=  \frac{1}{2\ell} \g_0 \tilde{Q}^{\pm}\,, &[P_1, \tilde{Q}^{\pm}] &=  \frac{1}{2\ell} \g_1 \tilde{Q}^{\mp} \,, \nn \\
			 [U, \tilde{Q}^{\pm}] &= \pm \frac{1}{2 \ell} \g_0 \tilde{Q}^{\pm} \,,
			\label{adspre1}
		\end{align}
		along with
		\bea
		\{ \tilde{Q}^+_\a ,  \tilde{Q}^+_\b  \} &=& - (\g_0 C^{-1})_{\a \b} P_0  +   (\g_0 C^{-1})_{\a \b} U \,,\nonumber\\
		\{ \tilde{Q}^+_\a ,  \tilde{Q}^-_\b  \} &=&  (\g_1 C^{-1})_{\a \b} P_1   +  \frac{1}{\ell}   (\g_* C^{-1})_{\a \b} J \,,\nonumber\\
		\{ \tilde{Q}^-_\a ,  \tilde{Q}^-_\b  \} &=&  - (\g_0 C^{-1})_{\a \b} P_0 -  (\g_0 C^{-1})_{\a \b} U \,.
		\label{adspre2}
		\eea
		Then, we consider the following expansion of the generators:
		\bea
		&& P_0 =  H + \eta^2 M \,, \quad  P_1 = \eta P\,, \quad  J= \eta G\,,  \quad U = U_1 + \eta^2 U_2\nn \\
		&& \tilde{Q}^+  = Q^+ + \eta^2 R\,, \quad \tilde{Q}^- = \eta Q^-\,, \label{expansions}
		\eea
		where $\eta$ is the expansion parameter.  In this way, we end up with the extended Newton-Hooke superalgebra sNH$_2$, which has the non-vanishing commutators\footnote{Closure of the algebras we introduce in the present paper has been verified also by means of the computer algebra program \textit{Cadabra} \cite{Peeters:2006kp,Peeters:2007wn}.}
		\begin{align}
			\left[ H, G\right]  &= -  P\,,          &  \left[G, P \right] &= M \,,          &  \left[ H, 
			P \right] &= \frac{1}{\ell^2} G\,, \nonumber\\
			[G, Q^+] &=  \frac{1}{2} \g_* Q^- \,, \nn &[G, Q^-] &=  \frac{1}{2} \g_* R \,, &[P, Q^+] &=  \frac{1}{2\ell} \g_1 Q^-\,, \nn \\
			[P, Q^-] &=  \frac{1}{2\ell} \g_1 R \,, & [H, Q^\pm] &=  \frac{1}{2\ell} \g_0 Q^\pm \,, &[H, R] &=  \frac{1}{2\ell} \g_0 R \,,  \nn \\
			[M, Q^+ ] &=  \frac{1}{2\ell} \g_0 R \,, & [U_1, Q^\pm] &= \pm \frac{1}{2 \ell}  \g_0 Q^\pm \,, &[U_1, R] &=  \frac{1}{2\ell}  \g_0 R\,, \nn \\
			[U_2, Q^+ ] &= \frac{1}{2\ell}  \g_0 R \,.
			\label{CosmoEBG1}
		\end{align}
		and anticommutators
		\bea
		\{ Q^+_\a ,  Q^+_\b  \} &=& - (\g_0 C^{-1})_{\a \b} H   +   (\g_0 C^{-1})_{\a \b} U_1  \,,\nonumber\\
		\{ Q^+_\a ,  Q^-_\b  \}  &=&  (\g_1 C^{-1})_{\a \b} P +  \frac{1}{\ell}   (\g_* C^{-1})_{\a \b} G\,,\nonumber\\
		\{ Q^+_\a ,  R_\b  \} &=& - (\g_0 C^{-1})_{\a \b} M   +   (\g_0 C^{-1})_{\a \b} U_2\,,\nn \\
		\{ Q^-_\a ,  Q^-_\b  \} &=& - (\g_0 C^{-1})_{\a \b} M   -  (\g_0 C^{-1})_{\a \b} U_2  \,.
		\label{NHN2}
		\eea
		The supersymmetric extended Newton-Hooke algebra above admits the invariant metric
		\bea
		&& \langle P,P \rangle = \frac{1}{\ell^2}\,, \quad \langle G,G \rangle =1\,, \quad \langle H,M \rangle = -\frac{1}{\ell^2}\,, \quad \langle U_1, U_2 \rangle = \frac{1}{\ell^2}\,,\nn\\
		&&  \langle Q^+_\alpha,R_\beta \rangle = \frac{2}{\ell}\left( C^{-1}\right)_{\a \b}\,,\quad \langle Q^-_\alpha,Q^-_\beta \rangle= \frac{2}{\ell} \left( C^{-1}\right)_{\a \b} \,.
		\label{MetricNH}
		\eea
		We can now move on to the construction of the non-relativistic JT supergravity action as a BF theory based on the sNH$_2$ structure obtained above. 
		
		\subsection{Non-relativistic Jackiw-Teitelboim supergravity action}
		
		Now that we have unveiled the sNH$_2$ structure, we can assign gauge fields for each Lie algebra generator, and the 1-form $A$ in the case at hand can be written as
		\bea
		A &=& \t H + e P + \o  G  + m M  +r_{1} U_1 + r_{2} U_2 + \bar\p^+ Q^+ + \bar\p^- Q^- + \bar\rho R \,.
		\label{GaugeField2}
		\eea
		The transformation rules for these gauge fields can be found by exploiting $\d_\epsilon A^I = d \e^I + C_{JK}{}^I \e^K A^J$, where $\e^A$ is the relevant gauge parameter and $C_{JK}{}^I$ are the structure constants, in this case, of sNH$_2$. 
		In particular, the (non-trivial) transformations along the parameters $\epsilon^{+\alpha}, \epsilon^{-\alpha}, \epsilon^\alpha$ (associated with $\psi^+,\psi^-,\rho$) are, respectively,  
	    \bea
		\d \t &=&   \bar{\e}^+ \g_0 \p^+ \,,\nn\\
		\d e &=&  - \bar{\e}^+ \g_{1} \p^- \,,\nn\\
		\d \o &=&  - \frac{1}{\ell} \bar{\e}^+ \g_* \p^- \,,\nn\\
		\d m &=&  \bar{\e}^+ \g_0 \rho \,,\nn\\
		\d r_{1}  &=&  -  \bar{\e}^+ \g_0 \p^+\,, \nn \\
		\d r_{2}  &=& - \bar{\e}^ + \g_0 \rho \,, \nn \\
		\d \p^+ &=& d  \e^+   +  \frac1{2\ell}  \g_0 \e^+ \t + \frac{1}{2\ell} \g_0 \e^+  r_{1} \,,\nn\\
		\d \p^- &=&  d \e^- + \frac12 \g_* \e^+ \o  +  \frac{1}{2\ell} \g_1 \e^+ e  \,,\nn\\
		\d \r &=&  \frac{1}{2\ell}   \g_0 \e^+ m +  \frac{1}{2\ell}   \g_0 \e^+ r_{2} \,,
		\label{NHsusytran+} 
		\eea
	    \bea
		\d e &=& - \bar{\e}^- \g_1 \p^+ \,,\nn\\
		\d \o &=&  - \frac{1}{\ell} \bar{\e}^- \g_* \p^+ \,,\nn\\
		\d m &=&  \bar{\e}^- \g_0 \p^- \,,\nn\\
		\d r_{2}  &=&   \bar{\e}^- \g_0 \p^-  \,, \nn \\
		\d \p^- &=&  d \e^- +  \frac{1}{2\ell}  \g_0 \e^- \t  -  \frac{1}{2\ell}   \g_0 \e^- r_{1} \,,\nn\\
		\d \r &=&   \frac12  \g_* \e^- \o +  \frac{1}{2\ell}  \g_1 \e^- e \,,
		\label{NHsusytran-} 
		\eea
	    \bea
		\d m &=& \bar{\epsilon} \g_0 \p^+  \,,\nn\\
		\d r_{2}  &=&  -  \bar{\epsilon} \g_0 \p^+ \,, \nn \\
		\d \r &=&  d  \epsilon  +   \frac{1}{2\ell} \g_0 \epsilon  \t + \frac{1}{2\ell}  \g_0 \epsilon  r_{1} \,.
		\label{NHsusytranrho} 
		\eea
		Besides, we have
		\bea
		\mathcal{X} &=& \Xi H + \Pi P + \Phi G + \Sigma M + T_1 U_1 + T_2 U_2 + \bar{\lambda}^+ Q^+ + \bar{\lambda}^- Q^- + \bar{\lambda} R \,,  \nn \\
		R&=& R(\tau) H + R(e) P + R(\omega) G + R(m) M + R(r_1) U_1 + R(r_2) U_2\nn \\
		&& + R(\psi^+) Q^+ + R(\psi^-) Q^- + R(\rho) R \,, \label{gaugeXandcurvNH}
		\eea
		where the sNH$_2$-covariant curvatures are given by
		\bea
		R (\tau) &=& d \t - \frac{1}{2} \bar\p^+  \g_0 \wedge \p^+   \,,\nn\\ 
		R(e) &=& d e  - \t  \wedge \o +  \bar\p^+  \g_1 \wedge \p^-  \,,\nn\\ 
		R(\omega) &=& d \o  + \frac{1}{\ell^2} \t \wedge e + \frac{1}{\ell}\bar\p^+  \g_* \wedge \p^- \,,\nn\\ 
		R (m) &=& d m + \o \wedge e -  \bar\p^+  \g_0 \wedge \r - \frac{1}{2} \bar\p^-  \g_0 \wedge \p^-\,, \nn \\
		R(r_1) &=& d r_1 +  \frac{1}{2} \bar\p^+  \g_0 \wedge \p^+\,, \nn \\
		R(r_2)&=& d r_2  + \bar\p^+  \g_0 \wedge \r - \frac{1}{2} \bar\p^-  \g_0 \wedge \p^-\,, \nn \\
		R (\psi^+) &=& d \p^+ + \frac{1}{2\ell}  \t \wedge \g_0 \p^+ + \frac{1}{2 \ell} r_1 \wedge \g_0 \p^+  \,,\nn\\ 
		R (\psi^-) &=& d  \p ^- + \frac{1}{2} \o \wedge \g_* \p^+     + \frac{1}{2\ell} \t \wedge \g_0 \p^-  + \frac{1}{2\ell} e \wedge \g_1 \p^+ -  \frac{1}{2\ell}  r_{1} \wedge \g_0 \p^-  \,,\nn\\
		R (\rho) &=& d\rho  +  \frac{1}{2}\o \wedge \g_* \p^- + \frac{1}{2\ell} \t \wedge \g_0 \rho  + \frac{1}{2 \ell} m \wedge \g_0 \p^+ +  \frac{1}{2\ell} e \wedge \g_1 \p^- \nn \\
		&& +  \frac{1}{2\ell} r_1 \wedge \g_0 \rho   + \frac{1}{2\ell} r_{2} \wedge \g_0 \p^+  \,. \label{susycurvatures}
		\eea
		As we have the invariant metric \eqref{MetricNH} and the curvatures \eqref{susycurvatures}, we are now able to write down the BF action for the non-relativistic JT supergravity theory, based on sNH$_2$, which reads
		\begin{align}
		S^{\text{NR}}_{\text{sJT}} &= \frac{k}{2\pi} \int \Big [\Phi R(\omega)  +  \frac{1}{\ell^2} \left( \Pi R(e) -  \Xi R(m) - \Sigma R(\tau)  + T_1 R(r_{2}) + T_2 R(r_{1})  \right)  \nn \\
		&+  \frac{2}{\ell}\left(\bar{\l}^+ R(\psi^+) +   \bar{\l}^- R(\psi^-) + \bar{\l} R(\rho)\right) \Big ]\,.\label{NHJTaction}
		\end{align}
		The action \eqref{NHJTaction} is a supersymmetric generalization of the bosonic one presented in \cite{Gomis2020,Grumiller2020}.  This non-relativistic JT supergravity action is invariant under the transformations \eqref{NHsusytran+}-\eqref{NHsusytranrho} together with the following (non-trivial) transformations of $\Phi, \Pi, \Xi, \Sigma, T_1, T_2, \lambda^+, \lambda^-, \lambda$, with parameters, in order of appearance, $\e^{+\alpha},\e^{-\alpha},\e$: 
		\bea
	\d \Phi &=& - \frac{1}{\ell} \bar{\e}^+  \g_* \l^-   \,, \nn\\
		\d \Pi &=& - \bar{\e}^+ \g_1 \l^-  \,, \nn\\
		\d \Xi &=&  \bar{\e}^+ \g_0 \l \,, \nn\\
		\d \Sigma  &=&  \bar{\e}^+ \g_0 \l^+  \,, \nn\\
		\d T_1 &=& - \bar{\e}^+ \g_0 \l \,, \nn\\
		\d T_2 &=& - \bar{\e}^+ \g_0 \l^+   \,,  \nn \\
		\d \l^+ &=&  \frac{1}{2\ell} \gamma_0 \Sigma \e^+ + \frac{1}{2\ell} \gamma_0 T_2 \e^+   \,, \nn \\
		\d \l^- &=& \frac{1}{2} \g_* \Phi \e^+ + \frac{1}{2 \ell} \g_1 \Pi \e^+ \,, \nn \\
		\d \l &=& \frac{1}{2\ell} \g_0 \Xi \e^+ + \frac{1}{2\ell} \gamma_0 T_1 \e^+ \,, \label{susylagtrans+}
		\eea
		\bea
	\d \Phi &=&  - \frac{1}{\ell} \bar{\e}^- \g_* \l  \,, \nn\\
		\d \Pi &=& - \bar{\e}^-   \g_1 \l \,, \nn\\
		\d \Sigma  &=&  \bar{\e}^- \g_0 \l^-  \,, \nn\\
		\d T_2 &=&  \bar{\e}^- \g_0 \l^-   \,,  \nn \\
		\d \l^+ &=&   \frac{1}{2} \gamma_\star \Phi \e^- + \frac{1}{2\ell} \g_1 \Pi \e^- \,, \nn \\
		\d \l^- &=&  \frac{1}{2\ell} \g_0 \Xi \e^- - \frac{1}{2\ell} \g_0 T_1 \e^- \,,  \label{susylagtrans-}
		\eea
		\bea
		\d \Sigma  &=&  \bar{\e} \g_0 \l \,, \nn\\
		\d T_2 &=&  -  \bar{\e} \g_0 \l \,,  \nn \\
		\d \l^+ &=&  \frac{1}{2\ell} \g_0 \Xi \e  +  \frac{1}{2\ell} \g_0 T_1 \e \,. \label{susylagtransrho}
		\eea
		Note that the latter can also be derived by exploiting the Lie algebra expansion method on \eqref{susytransformation}. 
		The equations of motion of the theory correspond to the vanishing of the sNH$_2$-covariant curvatures and coincide with the ones found in \cite{Gomis2020,Grumiller2020} when we restrict ourselves to the purely bosonic case. Notice that, in the supersymmetric theory, we end up with a non-vanishing torsion that is given, on-shell, in terms of a spinor bilinear.
		
		Finally, let us highlight that the same action \eqref{NHJTaction} can be obtained by directly performing the expansion procedure on the supersymmetric AdS$_2$ theory. To do so, one shall start by decomposing the AdS$_2$ curvatures \eqref{curvatures} as follows:
		\bea
		R(E)^0 &=& d E^0 +  \O \wedge E_1 -  \frac{1}{2} \delta_{ij}\bar{\Psi}^i \g_0 \wedge \Psi^j  \,, \nn\\
		R(E)^1 &=& d E^1 +  \O \wedge E^0  +  \frac{1}{2} \delta_{ij}\bar{\Psi}^i \g_1 \wedge \Psi^j  \,, \nn\\
		R(\Omega) &=& d\O  + \frac{1}{\ell^2}  E^0 \wedge E^1 + \frac{1}{2 \ell} \delta_{ij}\bar{\Psi}^i \g_* \wedge \Psi^j \,, \nn \\
		R(T) &=& d T  -  \frac{1}{2} \varepsilon_{ij} \bar{\Psi}^i  \wedge \Psi^j\,, \nn \\
		R(\Psi)^i &=& d \Psi ^i + \frac{1}{2} \O \wedge \g_* \Psi^i  + \frac{1}{2\ell} E^0 \wedge \g_0 \Psi^i + \frac{1}{2\ell} E^1 \wedge \g_1 \Psi^i  + \varepsilon^{ij} T \wedge  \Psi^j\,. \label{curvaturespreexpdec}
		\eea	
		Correspondingly, we can also implement the same decomposition on the action \eqref{sugraJT}, that is
		\bea
		\mathcal{S}_{\text{sJT dec.}} = \frac{k}{2\pi} \int \left[ \frac{1}{\ell^2}\left(X^0 R(E)_0 + X^1 R(E)_1  \right)+ X R(\O) + \frac{1}{\ell^2}Y  R(T) +  \frac{2}{\ell}\bar{\l}^i R(\Psi)_i \right] \,. \label{sJTpredec}
		\eea
		Subsequently, we introduce the spinors
	    \bea
		\Phi^\pm_\a = \frac{1}{\sqrt{2}} \left( \Psi^1_\a \pm (\g_0)_\a{}^\b \Psi^2_\b\right)\,. \label{fermionicdeftogetNHfromAdS}
		\eea
		Thus, the curvatures \eqref{curvaturespreexpdec} boil down to
        \bea
		R(E)^0 &=& d E^0 +  \O \wedge E_1 -  \frac{1}{2} \bar{\Phi}^+ \g_0 \wedge \Phi^+  -  \frac{1}{2} \bar{\Phi}^- \g_0 \wedge \Phi^-  \,, \nn\\
		R(E)^1 &=& d E^1 +  \O \wedge E^0  +  \bar{\Phi}^+ \g_1 \wedge \Phi^-  \,, \nn\\
		R(\Omega) &=& d\O  + \frac{1}{\ell^2}  E^0 \wedge E^1 + \frac{1}{\ell} \bar{\Phi}^+ \g_* \wedge \Phi^- \,, \nn \\
		R(T) &=& d T  +  \frac{1}{2} \bar{\Phi}^+  \wedge \Phi^+ -  \frac{1}{2} \bar{\Phi}^-  \wedge \Phi^-\,, \nn \\
		R(\Phi^\pm) &=& d \Phi ^\pm + \frac{1}{2} \O \wedge \g_* \Phi^\mp  + \frac{1}{2\ell} E^0 \wedge \g_0 \Phi^\pm + \frac{1}{2\ell} E^1 \wedge \g_1 \Phi^\mp  \pm \frac{1}{2\ell} T \wedge \g_0  \Phi^\pm\,. \label{curvaturesNHpreexpspinors}
		\eea	
		We can then see that, performing the expansion
		\bea
		&& E^0 =  \t + \eta^2 m \,, \quad  E^1 = \eta e \,, \quad  \O= \eta \o\,,  \quad T = r_1 + \eta^2 r_2 \,, \nn \\
		&& \Phi^+  = \psi^+ + \eta^2 \r \,, \quad \Phi^- = \eta \psi^- \,, \label{expansionstogetNH}
		\eea
		we precisely end up with the sNH$_2$ curvatures \eqref{susycurvatures}. 
		Finally, the expansion of the action yields
	\bea
	 \overset{(2)}{\mathcal{S}}_{\text{NR sJT}} &=& \frac{k}{2\pi} \int \Bigg[ \frac{1}{\ell^2}\left(\overset{(2)}{X_0} R(\tau)+\overset{(0)}{X_0} R(m)  + \overset{(1)}{X_1} R(e)  \right)+ \overset{(1)}{X} R(\omega)  + \frac{1}{\ell^2}\left(\overset{(2)}{Y} R(r_1) + \overset{(0)}{Y} R(r_2)\right) \nn \\
	 &&+\frac{2}{\ell} \left( \overset{(2)}{\bar{\l}^+} R(\psi^+)  + \overset{(0)}{\bar{\l}} R(\rho)  + +\overset{(1)}{\bar{\l}^-} R(\psi^-)\right)    \Bigg]\,,
	\eea 
	that is, with the identifications 
	\bea
	&&\overset{(2)}{X_0} = -\Sigma\,, \quad \overset{(0)}{X_0} = - \Xi\,, \quad \overset{(1)}{X_1} = \Pi\,, \quad \overset{(1)}{X} = \Phi\,, \quad \overset{(2)}{Y} = T_2\,, \quad \overset{(0)}{Y} = T_1\,, \nn \\
	&&\overset{(2)}{\bar{\l}^+} = \bar{\l}^+\,, \quad  \overset{(0)}{\bar{\l}} = \bar{\l}\,, \quad \overset{(1)}{\bar{\l}^-} = \bar{\l}^-\,,
	\eea 
	the non-relativistic JT supergravity action \eqref{NHJTaction}.

		\section{Ultra-relativistic Jackiw-Teitelboim supergravity}\label{URthy}
		
		In this section we will introduce a supersymmetric extension of the extended AdS$_2$ Carroll algebra which will allow us to develop the corresponding Carrollian, that is ultra-relativistic, JT supergravity action as a metric BF theory.
		
		\subsection{Supersymmetric extended AdS$_2$ Carroll algebra as a redefinition of sNH$_2$}
		
		The purely bosonic extended AdS$_2$ Carroll structure first appeared in \cite{Grumiller2020} and \cite{Gomis2020}. We shall now present a supersymmetric extension of the latter as a redefinition of sNH$_2$.
		To this aim, let us therefore perform the following redefinition on the sNH$_2$ structure previously obtained:
		\begin{equation}
		    H \leftrightarrow P \,, \quad \ell \rightarrow - i \ell \,, \quad \gamma_0 \rightarrow - i \gamma_1 \,, \quad \gamma_1 \rightarrow - i\gamma_0 \,, \quad \gamma_\star \rightarrow \gamma_\star \,, \quad C \rightarrow - i C \,. \label{map}
		\end{equation}
		By doing so, we end up with the supersymmetric extended AdS$_2$ Carroll algebra, which reads
		\begin{align}
		\left[ H, G\right]  &= -  M\,,          &  \left[G, P \right] &= H \,,          &  \left[ H, P \right] &= \frac{1}{\ell^2} G\,, \nonumber\\
		[G, Q^+] &=  \frac{1}{2} \g_* Q^- \,, \nn &[G, Q^-] &=  \frac{1}{2} \g_* R \,, &[H, Q^+] &=  \frac{1}{2\ell} \g_0 Q^-\,, \nn \\
		[H, Q^-] &=   \frac{1}{2\ell} \g_0 R \,, & [P, Q^\pm] &=  \frac{1}{2\ell} \g_1 Q^\pm \,, &[P, R] &=  \frac{1}{2\ell} \g_1 R \,,  \nn \\
		[M, Q^+ ] &=   \frac{1}{2\ell} \g_1 R \,, & [U_1, Q^\pm] &= \pm \frac{1}{2 \ell}  \g_1 Q^\pm \,, &[U_1, R] &=   \frac{1}{2\ell}  \g_1 R\,, \nn \\
		[U_2, Q^+ ] &=  \frac{1}{2\ell}  \g_1 R \,, \nn \\
		\label{carrollsusycomm}
	\end{align}
	together with
	\bea
	\{ Q^+_\a ,  Q^+_\b  \} &=& (\g_1 C^{-1})_{\a \b} P   -   (\g_1 C^{-1})_{\a \b} U_1  \,,\nonumber\\
	\{ Q^+_\a ,  Q^-_\b  \}  &=&  - (\g_0 C^{-1})_{\a \b} H + \frac{1}{\ell}(\g_* C^{-1})_{\a \b} G\,,\nonumber\\
	\{ Q^+_\a ,  R_\b  \} &=&  (\g_1 C^{-1})_{\a \b} M   -  (\g_1 C^{-1})_{\a \b} U_2\,,\nn \\
	\{ Q^-_\a ,  Q^-_\b  \} &=&  (\g_1 C^{-1})_{\a \b} M   +  (\g_1 C^{-1})_{\a \b} U_2  \,.
	\label{carrollsusyanticomm}
	\eea
	Note that this construction is reminiscent of the duality existing at the purely bosonic level in two spacetime dimensions between the extended NH$^\pm$ and the extended (A)dS$_2$ Carroll algebras.
	
	Let us mention that the same extended AdS$_2$ Carroll superalgebra \eqref{carrollsusycomm}-\eqref{carrollsusyanticomm} can also be obtained from the $\mathcal{N}=2$ AdS$_2$ superalgebra \eqref{ads2susy1}-\eqref{ads2susy2}, once decomposed as in \eqref{adsdecom1}-\eqref{adsdecom2}, by introducing the fermionic charges
	\bea
		F^{\pm} = \frac{1}{\sqrt{2}} \left(Q^1 \pm  i \g_1 Q^2\right)
	\eea 
	and performing the redefinition $U \rightarrow i U$ with the subsequent expansion
	\bea
	    &&P_0 =  \eta H  \,, \quad  P_1 = P + \eta^2 M\,, \quad  J= \eta G\,,  \quad U = U_1 + \eta^2 U_2\nn \\
		&&F^+  = Q^+ + \eta^2 R\,, \quad F^- = \eta Q^-\,. \label{expansions2}
	\eea
	
	The extended AdS$_2$ Carroll superalgebra \eqref{carrollsusycomm}-\eqref{carrollsusyanticomm} is endowed with the invariant metric
	\bea
		&& \langle H,H \rangle = -\frac{1}{\ell^2}\,, \quad \langle G,G \rangle =1\,, \quad \langle P,M \rangle = \frac{1}{\ell^2}\,, \quad \langle U_1, U_2 \rangle = - \frac{1}{\ell^2}\,,\nn\\
		&&  \langle Q^+_\alpha,R_\beta \rangle = \frac{2}{\ell}\left( C^{-1}\right)_{\a \b}\,,\quad \langle Q^-_\alpha,Q^-_\beta \rangle = \frac{2}{\ell} \left( C^{-1}\right)_{\a \b} \,,
		\label{MetricAdSC}
    \eea
	which, in particular, can be obtained from \eqref{MetricNH} by means of the mapping in \eqref{map}.

		\subsection{Carrollian Jackiw-Teitelboim supergravity}
	
	    We shall now proceed with the development of the ultra-relativistic BF JT supergravity theory based on the Carrollian superalgebra just unveiled. The 1-form $A$ has the same form of \eqref{GaugeField2}, and also the implicit form of $\mathcal{X}$ and $R$ is formally the same as in \eqref{gaugeXandcurvNH}.\footnote{This is just matter of our notation, in fact. We make a little abuse of notation in order to avoid the introduction of new symbols here.} On the other hand, now the (non-trivial) transformations along the parameters $\e^{+\alpha}, \e^{-\alpha}, \e^\alpha$ are respectively given by
	\bea
	\d \t &=&    \bar{\e}^+ \g_{0} \p^-   \,,\nn\\
    \d e &=&  - \bar{\e}^+ \g_1 \p^+ \,,\nn\\
    \d \o &=& -  \frac{1}{\ell} \bar{\e}^+ \g_* \p^-  \,,\nn\\
    \d m &=&  -  \bar{\e}^+ \g_1 \rho  \,,\nn\\
    \d r_{1}  &=&   \bar{\e}^+ \g_1 \p^+\,, \nn \\
    \d r_{2}  &=&  \bar{\e}^+ \g_1 \rho \,, \nn \\
    \d \p^+ &=& d  \e^+   +  \frac1{2\ell} \g_1 \e^+ e + \frac{1}{2\ell}  \g_1 \e^+ r_{1} \,,\nn\\
    \d \p^- &=&  \frac12  \g_* \e^+ \o +  \frac{1}{2\ell}  \g_0 \e^+ \t  \,,\nn\\
    \d \r &=&  \frac{1}{2\ell}  \g_1 \e^+   m   +  \frac{1}{2\ell} \g_1 \e^+ r_{2} \,,
    \label{Carsusytran+} 
    \eea
    \bea
	\d \t &=&    \bar{\e}^- \g_0 \p^+ \,,\nn\\
    \d \o &=& -  \frac{1}{\ell} \bar{\e}^- \g_* \p^+ \,,\nn\\
    \d m &=& - \bar{\e}^- \g_1 \p^-  \,,\nn\\
    \d r_{2}  &=&  -  \bar{\e}^- \g_1 \p^-  \,, \nn \\
    \d \p^- &=&  d \e^-  +  \frac{1}{2\ell}  \g_1 \e^- e  -  \frac{1}{2\ell}  \g_1 \e^- r_{1} \,,\nn\\
    \d \r &=&    \frac12 \g_* \e^-  \o  +   \frac{1}{2\ell}  \g_0 \e^- \t \,,
    \label{Carsusytran-} 
    \eea
	\bea
    \d m &=&  -  \bar{\e} \g_1 \p^+  \,,\nn\\
    \d r_{2}  &=&  \bar{\e} \g_1 \p^+ \,, \nn \\
    \d \r &=&  d  \e  \frac{1}{2\ell}  \g_1 \e e +   \frac{1}{2\ell}   \g_1 \e r_{1} \,.
    \label{Carsusytranrho} 
    \eea	
	The extended super-AdS$_2$ Carroll-covariant curvatures read	
		\bea
        R (\tau) &=& d \t  + \o \wedge \e  - \bar\p^+ \g_0 \wedge \p^-  \,,\nn\\ 
        R(e) &=& d e    +    \frac{1}{2} \bar\p^+ \g_1 \wedge \p^+  \,,\nn\\ 
        R(\omega) &=& d \o  + \frac{1}{\ell^2} \t \wedge e  + \frac{1}{\ell}\bar\p^+ \g_* \wedge \p^- \,,\nn\\ 
        R (m) &=& d m + \o \wedge \t+  \bar\p^+ \g_1 \wedge \r + \frac{1}{2} \bar\p^- \g_1 \wedge \p^-\,, \nn \\
        R(r_1) &=& d r_1 -  \frac{1}{2} \bar\p^+ \g_1 \wedge \p^+\,, \nn \\
        R(r_2)&=& d r_2  -  \bar\p^+ \g_0 \wedge \r +  \frac{1}{2} \bar\p^- \g_1 \wedge \p^-\,, \nn \\
        R (\psi^+) &=& d \p^+   + \frac{1}{2\ell}  e \wedge \g_1 \p^+ +   \frac{1}{2 \ell} r_1 \wedge \g_{1} \p^+  \,,\nn\\ 
        R (\psi^-) &=& d  \p ^- + \frac{1}{2} \o \wedge \g_* \p^+     + \frac{1}{2\ell} \t \wedge \g_0 \p^+  + \frac{1}{2\ell} e \wedge \g_1 \p^- -  \frac{1}{2\ell}  r_{1} \wedge \g_1 \p^-  \,,\nn\\
        R (\rho) &=& d\rho  +  \frac{1}{2}\o \wedge \g_* \p^- + \frac{1}{2\ell} \t \wedge \g_0 \psi^-  + \frac{1}{2 \ell} m \wedge \g_1 \p^+ +  \frac{1}{2\ell} e \wedge \g_1 \r  \nn\\
        && +  \frac{1}{2\ell} r_1 \wedge \g_1 \rho   +  \frac{1}{2\ell} r_{2} \wedge \g_1 \p^+  \,. \label{Carsusycurvatures}
        \eea
        Therefore, using the invariant metric \eqref{MetricAdSC} together with the curvatures \eqref{Carsusycurvatures}, we end up with the following ultra-relativistic JT supergravity action as a metric BF theory based on the extended AdS$_2$ Carroll superalgebra \eqref{carrollsusycomm}-\eqref{carrollsusyanticomm}:
	\begin{align}
	\mathcal{S}^{\text{UR}}_{\text{sJT}} &= \frac{k}{2\pi} \int \Big [\Phi R(\omega)  +  \frac{1}{\ell^2} \left( \Sigma R(e) + \Pi R(m) - \Xi R(\tau)  - T_1 R(r_2)-  T_2 R(r_1) \right)  \nn \\
	&+ \frac{2}{\ell} \left( \bar{\l}^+ R(\psi^+) +   \bar{\l}^- R(\psi^-) + \bar{\l} R(\rho) \right) \Big ]\,.\label{CJTaction}
    \end{align}
    The ultra-relativistic action \eqref{CJTaction} is a supersymmetric generalization of the purely bosonic Carrollian one in \cite{Gomis2020,Grumiller2020}. It is invariant under the transformations \eqref{Carsusytran+}-\eqref{Carsusytranrho} together with the following (non-trivial) transformations of the fields $\Phi, \Sigma, \Xi, \Pi, T_1, T_2, \lambda^+, \lambda^-, \lambda$, along the parameters $\e^{+\alpha},\e^{-\alpha},\e$, in order of appearance:
		\bea
		\d \Phi &=&  - \frac{1}{\ell} \bar{\e}^+ \g_* \l^-   \,, \nn\\
		\d \Sigma &=& - \bar{\e}^+  \g_1 \l^+  \,, \nn\\
		\d \Xi &=&  \bar{\e}^+ \g_0 \l^-  \,, \nn\\
		\d \Pi  &=&  - \bar{\e}^+ \g_1 \l  \,, \nn\\
		\d T_1 &=& \bar{\e}^+ \g_1 \l   \,, \nn\\
		\d T_2 &=&  \bar{\e}^+ \g_1 \l^+    \,,  \nn \\
		\d \l^+ &=& \frac{1}{2\ell} \g_1 \Sigma \e^+ + \frac{1}{2\ell} \g_1 T_2 \e^+ \,, \nn\\
		\d \l^- &=& \frac{1}{2} \g_* \Phi \e^+ + \frac{1}{2\ell} \g_0 \Xi \e^+  \,, \nn\\
		\d \l &=&  \frac{1}{2\ell} \g_1 \Pi \e^+ + \frac{1}{2\ell} \g_1 T_1 \e^+ \,,\label{carsusylag+}
		\eea
		\bea
		\d \Phi &=&  -  \frac{1}{\ell} \bar{\e}^- \g_* \l \,, \nn\\
		\d \Sigma &=& -  \bar{\e}^-  \g_1 \l^- \,, \nn\\
		\d \Xi &=&   \bar{\e}^- \g_0 \l \,, \nn\\
		\d T_2 &=&  -  \bar{\e}^- \g_1 \l^- \,,  \nn \\
		\d \l^+ &=&  \frac{1}{2} \gamma_\star \Phi \e^- + \frac{1}{2\ell} \g_1 \Xi \e^- \,, \nn\\
		\d \l^- &=& \frac{1}{2\ell} \g_1 \Pi \e^- - \frac{1}{2\ell} \g_1 T_1 \e^- \,,  \label{carsusylag-}
		\eea
		\bea
		\d \Sigma &=&  -  \bar{\e} \g_1 \l \,, \nn\\
		\d T_2 &=&  \bar{\e} \g_1 \l  \,,  \nn \\
		\d \l^+ &=&  \frac{1}{2\ell} \g_1 \Pi \e + \frac{1}{2\ell} \g_1 T_1 \e \,.\label{carsusylagrho}
		\eea
        The equations of motion of this Carrollian theory correspond to the vanishing of the curvatures \eqref{Carsusycurvatures}, and they boil down to the ones obtained in \cite{Gomis2020,Grumiller2020} if we restrict ourselves to the purely bosonic case. In the Carrollian supersymmetric theory above we end up with an on-shell non-vanishing torsion given in terms of a fermion bilinear.
		
		Let us conclude by observing that the same ultra-relativistic action \eqref{CJTaction} can be derived by expanding the ${\mathcal{N}}=2$ super-AdS$_2$ relativistic one. More precisely, starting from the decomposed expressions \eqref{curvaturespreexpdec} and \eqref{sJTpredec} and introducing the (new) spinors
		\bea
		\Phi^\pm_\a = \frac{1}{\sqrt{2}} \left( \Psi^1_\a \pm i(\g_1)_\a{}^\b \Psi^2_\b\right) \,, \label{fermionicdefpreexptoCarroll}
		\eea
		we get the following ``pre-expanded'' curvatures:
        \bea
		R(E)^0 &=& d E^0 +  \O \wedge E_1 -   \bar{\Phi}^+ \g_0 \wedge \Phi^-\,, \nn\\
		R(E)^1 &=& d E^1 +  \O \wedge E^0  + \frac{1}{2} \bar{\Phi}^+ \g_1 \wedge \Phi^+  + \frac{1}{2} \bar{\Phi}^- \g_1 \wedge \Phi^-  \,, \nn\\
		R(\Omega) &=& d\O  + \frac{1}{\ell^2}  E^0 \wedge E^1 + \frac{1}{\ell} \bar{\Phi}^+ \g_* \wedge \Phi^- \,, \nn \\
		R(T) &=& d T  -  \frac{1}{2}  \bar{\Phi}^+  \wedge \Phi^+ +  \frac{1}{2}  \bar{\Phi}^-  \wedge \Phi^-\,, \nn \\
		R(\Phi^\pm) &=& d \Phi ^\pm + \frac{1}{2} \O \wedge \g_* \Phi^\mp  + \frac{1}{2\ell} E^0 \wedge \g_0 \Phi^\mp + \frac{1}{2\ell} E^1 \wedge \g_1 \Phi^\pm  \pm \frac{1}{2\ell}  T \wedge \g_1  \Phi^\pm\,. \label{curvaturespreexptoCarroll}
		\eea	
		Then, performing the expansion
		\bea
		&& E^0 =  \eta \t \,, \quad  E^1 = e + \eta^2 m \,, \quad  \O= \eta \o \,,  \quad T = r_1 + \eta^2 r_2 \,, \nn \\
		&& \Phi^+  = \psi^+ + \eta^2 \r\,, \quad \Phi^- = \eta \psi^- \,, \label{expansionstoCarroll}
		\eea
		we recover the extended super-AdS$_2$ Carroll-covariant curvatures \eqref{Carsusycurvatures}. Finally, on the same lines of what we have done in the non-relativistic case, expanding the action \eqref{sJTpredec} we find
		\bea
	    \overset{(2)}{\mathcal{S}}_{\text{UR sJT}} &=& \frac{k}{2\pi} \int \Bigg[ \frac{1}{\ell^2}\left(\overset{(1)}{X_0} R(\tau)+ \overset{(0)}{X_1} R(m)  + \overset{(2)}{X_1} R(e)  \right)+ \overset{(1)}{X} R(\omega)  + \frac{1}{\ell^2}\left(\overset{(2)}{Y} R(r_1) + \overset{(0)}{Y} R(r_2)\right) \nn \\
	    &&+\frac{2}{\ell} \left( \overset{(2)}{\bar{\l}^+} R(\psi^+)  + \overset{(0)}{\bar{\l}} R(\rho)  + +\overset{(1)}{\bar{\l}^-} R(\psi^-)\right)    \Bigg] \,.
	    \eea 
		Finally, with the identifications
	    \bea
    	&&\overset{(1)}{X_0} = -\Xi\,, \quad \overset{(2)}{X_1} =  \Sigma\,, \quad \overset{(0)}{X_1} = \Pi\,, \quad \overset{(1)}{X} = \Phi\,, \quad \overset{(2)}{Y} = - T_2\,, \quad \overset{(0)}{Y} = - T_1\,, \nn \\
    	&&\overset{(2)}{\bar{\l}^+} = \bar{\l}^+\,, \quad  \overset{(0)}{\bar{\l}} = \bar{\l}\,, \quad \overset{(1)}{\bar{\l}^-} = \bar{\l}^-\,,
    	\eea 
		we recover exactly the Carrollian JT supergravity action \eqref{CJTaction}.

		\section{Conclusions}\label{concl}
		
		In this paper, we have presented supersymmetric extensions of non- and ultra-relativistic JT gravity. In particular, starting from the fact that $\mathcal{N}=2$ JT supergravity can be formulated at first-order as a metric BF theory based on the $\mathcal{N}=2$ AdS$_2$ superalgebra, we have exploited the Lie algebra expansion method to develop its non-relativistic counterpart, which has its roots in the superalgebra we named sNH$_2$. 
		After that, by redefining some quantities appearing in sNH$_2$, we have obtained the extended AdS$_2$ Carroll superalgebra, together with the associated invariant metric. The latter has then be used to write down the Carrollian JT supergravity action as a metric BF theory. 
		Remarkably, the mapping from sNH$_2$ to the supersymmetric extended AdS$_2$ Carroll algebra can be seen as a supersymmetric extension of the duality ``extended NH$^\pm$ $\leftrightarrow$ extended (A)dS Carroll'' existing at the purely bosonic level in two dimensions. 
		Furthermore, we have explicitly shown that the same non-relativistic and ultra-relativistic JT supergravity theories presented here can also be obtained by directly applying the expansion procedure on the $\mathcal{N}=2$ JT supergravity action.
		
		{Observe that, in the non-relativistic case, the field equations (correspinding to the vanishing of the non-relativistic supercurvatures) allow to completely solve the non-relativistic spin connection as (see also \cite{Gomis2020})
        \begin{equation}\label{omegaNR}
        \omega^{\text{NR}}_\mu = 2 \tau^{[\alpha} e^{\beta]} \left( e_\mu \partial_\alpha e_\beta - \tau_\mu \partial_\alpha m_\beta \right) + \text{fermion bilinears} \,.
        \end{equation}
        Analogously, in the ultra-relativistic theory, the ultra-relativistic spin connection can be entirely solved as (cf. also \cite{Gomis2020})
        \begin{equation}\label{omegaUR}
        \omega^{\text{UR}}_\mu = 2 e^{[\alpha} \tau^{\beta]} \left( \tau_\mu \partial_\alpha \tau_\beta - e_\mu \partial_\alpha m_\beta \right) + \text{fermion bilinears} \,.
        \end{equation}
        Let us also mention that, unlike the relativistic case, by plugging the expressions \eqref{omegaNR} and \eqref{omegaUR} back into the first-order non-relativistic and ultra-relativistic actions, respectively, one obtains a theory that, in general, is not dynamically equivalent to its first-order counterpart. However, the equivalence is obtained by considering the non-relativistic and ultra-relativistic sectors $d\tau=\frac{1}{2}\bar{\psi}^+ \gamma_0 \wedge \psi^+$ and $de=-\frac{1}{2} \bar{\psi}^+ \gamma_1 \wedge \psi^+$, respectively.}
		
		The expansions we have performed in this work also enables us to consider BF theories beyond the supersymmetric Newton-Hooke and AdS Carrollian ones, and this would be particularly interesting from the Post-Newtonian expansion and \textit{large c} expansion point of view \cite{Gomis2020,VandenBleeken,Hansen2020}. Furthermore, the procedure we have presented in this paper could be useful to obtain the next order boundary Schwarzian actions (which are dual boundary actions of next order gravity actions) by expanding the related Maurer-Cartan forms. 
		
		Another possible future direction consists in considering supersymmetric extensions of the Carrollian and non-relativistic boundary Schwarzian actions in \cite{Grumiller2020, Gomis2020} along the lines of \cite{Kozyrev2021} (some work is currently in progress on this point). Besides, the non-relativistic JT supergravity theory may serve as a starting point to consider the supersymmetric extension of the flat space boundary action that appeared in \cite{Afshar2019}. It would also be interesting to consider ultra- and non-relativistic limits (and associated supersymmetric extensions) of the most general deformation of JT gravity studied in \cite{Grumiller:2021cwg,Grumiller:2002nm}.

		\section*{Acknowledgements}
		
		We thank Mehmet Ozkan and Patricio Salgado-Rebolledo for useful comments. 
		L.R. would like to thank the Department of Applied Science and Technology of the Polytechnic of Turin for financial support. U.Z. is supported by TUBITAK - 2218  National Postdoctoral Research Fellowship Program  under grant number 118C512. 
		
		\appendix
		
		\section{Notation and conventions}\label{appA}
		
		The two-dimensional spacetime metric is defined as $\eta_{AB}=\diag(-,+)$. In our notation, $\ell$ denotes the AdS$_2$ radius and the cosmological constant is $\Lambda = - \frac{1}{\ell^2}$. Regarding gamma matrices, we have $\gamma^A = (\gamma^0 , \gamma^1)$, with
		\begin{equation}
		    \gamma^0 = - \gamma_0 = \begin{pmatrix}
		    0 & 1 \\
		    -1 & 0 
		    \end{pmatrix} = i \sigma_2 \,, \quad \gamma^1 = \gamma_1 = \begin{pmatrix}
		    0 & 1 \\
		    1 & 0 
		    \end{pmatrix} = \sigma_1 \,, \quad
		    \gamma_\star = - \gamma_0 \gamma_1 = \begin{pmatrix}
		     1 & 0 \\
		    0 &  - 1 
		    \end{pmatrix} = \sigma_3 \,.
		\end{equation}
		The charge conjugation matrix is defined as
		\begin{equation}
		    C = i \sigma_2 = \begin{pmatrix}
		    0 & 1 \\
		    -1 & 0 
		    \end{pmatrix} \,.
		\end{equation}


\end{document}